\documentstyle[twocolumn,prl,aps]{revtex}
\begin{document}
\title{Anomalous spin-splitting of two-dimensional electrons in an AlAs Quantum
Well}
\draft
\author{S. J. Papadakis, E. P. De Poortere, and M. Shayegan}
\address{Department of Electrical Engineering, Princeton University, Princeton,
NJ  08544}
\date{August 1998}
\maketitle
\begin{abstract}
We measure the effective Land\'e {\it g}-factor of high-mobility two-dimensional
electrons in a modulation-doped AlAs quantum well by tilting the sample in a
magnetic field and monitoring the evolution of the magnetoresistance
oscillations.  The data reveal that $|g| = 9.0$, which is much enhanced with
respect
to the reported bulk value of 1.9.  Surprisingly, in a large range of magnetic
field and Landau level fillings, the value of the enhanced {\it g}-factor
appears to be constant.
\end{abstract}
\pacs{73.40.Kp, 73.40.-c, 72.15.Gd, 73.40.Hm}

The effective Land\'e {\it g}-factor and effective mass {$m^*$} are two
fundamental parameters that characterize the energy levels of two-dimensional
electron systems (2DESs) in semiconductors in the presence of a
magnetic field ($B$).  In a simple, non-interacting picture, the cyclotron
energy ($\hbar\omega_c \equiv {\hbar}eB_{\perp}/m^*$)  associated with the
electron's
orbital motion determines the separation between the quantized energy levels
(Landau levels), while the Zeeman energy ($g{\mu_B}B$) gives the
"spin-splitting" of the Landau levels ($B_{\perp}$ is the component of $B$
perpendicular to the 2DES plane).

For 2DESs in a high $B_{\perp}$ it is well known that when there
are unequal populations of electrons with opposite spin, electron-electron
interaction can lead to a substantial enhancement of the spin-splitting energy
which can in turn be expressed as an enhancement of the effective {\it
g}-factor \cite{Fang68,Janak69,Ando74}.  In GaAs 2DESs, for example, the
exchange enhancement of the {\it g}-factor leads to the energy gaps for the
quantum Hall effect states at odd Landau level fillings ($\nu$) being much
larger than the bare Zeeman energy \cite{Nicholas88}.  Moreover, the magnitude
of the {\it g}-factor enhancement oscillates with $\nu$ as the spin population
difference does \cite{Ando74,Nicholas88,Englert82,Nicholas83,Weitz96,Shen96}.

We report here an experimental determination of the spin-splitting energy for
electrons confined to a modulation-doped AlAs quantum well (QW).  In contrast to
GaAs, where electrons occupy the conduction band minimum at the Brillouin zone
center ($\Gamma$-point) and form a spherical Fermi surface, in AlAs they occupy
conduction band ellipsoids near the zone edge ($X$-point).  This is somewhat
similar to the case of 2D electrons at the Si/SiO$_2$ (100) interface except
that in the AlAs QW that we have studied, an ellipsoid with its major
axis parallel (as opposed to perpendicular) to the 2D plane is
occupied \cite{Papadakis98}.  In our
measurements we utilize the "coincidence" method, a technique which has been
used to study the {\it g}-factor enhancement in other 2DESs such as those in
Si/SiO$_2$ \cite{Fang68}, SiGe \cite{Weitz96}, and GaAs \cite{Nicholas88}.  The
results are surprisingly simple yet puzzling:  in a large range of $\nu$, we
find a significant enhancement of the
{\it g}-factor with respect to the reported bulk value but, remarkably, the
enhancement appears to be independent of $\nu$.  The 2DES behaves like a
non-interacting system of electrons but with a much-enhanced {\it g}-factor.

The experiment was done on samples from two wafers that were grown by molecular
beam epitaxy on undoped GaAs (100) substrates.  In both wafers the 2DES is
confined to a 150 \AA-wide AlAs QW which is separated from the Si dopants by
AlGaAs barriers.  Three samples (A, B, and C) from wafer 1 and one sample (D)
from wafer 2 were used in the tilt experiment.  Sample A was
photolithographically patterned with an L-shaped Hall
bar whose two perpendicular arms lay on the [100] and [010] directions.
Samples B, C, and D had a van der Pauw geometry.  Samples A and B had evaporated
metal front gates to control the density.  The experiments were performed in a
pumped $^3$He system at a temperature of 0.3 K, in magnetic fields up to 16 T.
The samples were mounted on a platform which could be rotated {\it in situ}.
The ungated carrier density of sample A was $n = 2.08 \times
10^{11}$ cm$^{-2}$ and the mobilities along the two arms of the L-shaped hall
bar were 6.1 m$^{2}$/Vs for the high-mobility direction and 4.2
m$^{2}$/Vs for the low-mobility direction \cite{Papadakis98}.

The effective masses for the conduction band ellipsoids in bulk AlAs are $m_l =
1.1m_e$ for the longitudinal mass and $m_t = 0.19m_e$ for the transverse
mass \cite{Adachi85}.  For QWs of width greater than $\sim$60 {\AA,} the
2D electrons will be forced to occupy the two ellipsoids whose major axes lie in
the plane of the 2DES \cite{vandeStadt95,vanKesteren89,Maezawa91}.  In our
samples, measurements have shown that only one of the two in-plane ellipsoids is
occupied \cite{Papadakis98,Lay93}.  In particular, cyclotron resonance
measurements reveal a cyclotron resonance effective mass of $m_{CR} =
0.46m_e$, in excellent agreement with the mass, $\sqrt{m_lm_t}$, expected for
in-plane ellipsoids \cite{Lay93}.  This observation is consistent with the work
of Smith {\it et al.}, who also conclude that in multiple AlAs QW
samples with a QW width of 150 {\AA} only a single in-plane
ellipsoid with similar $m_{CR}$ is occupied \cite{TPSmith87}.

We used the coincidence method \cite{Fang68} to determine
the product of the Land\'e {\it g}-factor and the effective mass ($|gm^*|$) of
the electrons in the AlAs QW.  Note that this method cannot determine the sign
of $g$.  When a 2DES is
tilted in a magnetic field, the Zeeman energy $g{\mu_B}B$ changes relative to
the cyclotron energy $\hbar\omega_c$ because the Zeeman energy is proportional
to the total $B$ while
the Landau level separation depends on $B_{\perp}$.  At the coincidence angles,
spin-up and spin-down levels of different Landau levels become degenerate.  This
degeneracy can be seen in
magnetoresistance data.  At a coincidence angle, in an ideal non-interacting
system, half of the longitudinal resistance ($R_{xx}$) minima, corresponding to
half of the integer $\nu$ (either the even or the odd), disappear.  The other
half reach a maximum strength.  Once the angle at which a coincidence occurs is
found, $|gm^*|$ can be determined from the equation
\begin{equation}
l\hbar\omega_c = |g|{\mu_B}B \;,
\end{equation}
where $l$ is an index determined by both the relative values of $|g|{\mu_B}B$
and $\hbar\omega_c$ at $\theta = 0$ and the order of the coincidence observed.
For
example, if $|g|{\mu_B}B = 0.3\hbar\omega_c$ at $\theta = 0$, then at the first
coincidence angle ($\theta_1$) $l = 1$, at the second coincidence angle
($\theta_2$) $l = 2$, etc.  However, if $|g|{\mu_B}B = 1.3\hbar\omega_c$ at
$\theta = 0$, then for $\theta_1$, $l = 2$; for $\theta_2$, $l = 3$; and so on.
For all of the coincidence measurements in other materials that we cite, $l = 1$
for $\theta_1$; i.e. the Zeeman energy is {\it smaller} than the cyclotron
energy at $\theta =
0$ \cite{Ando74,Nicholas88,Englert82,Nicholas83,Weitz96,Shen96}.  Our data
reveals that the opposite is true for the 2DES in AlAs QWs that we have
studied.

Experimentally, we first made sure
that the sample was at zero angle ($\theta = 0$) by maximizing the Hall
resistance in a small $B$.  Then we made magnetoresistance measurements at
various $\theta$, determining $\theta$ by comparing the Hall resistances and the
positions of the $R_{xx}$ minima to those of the $\theta = 0$ trace.  Data from
sample A, at a density of $1.4 \times 10^{11}$ cm$^{-2}$, are shown in Fig.
\ref{lown}.  $R_{xx}$ traces for various
angles, offset vertically for clarity, are plotted vs. $B_{\perp}$.
Concentrating on $\nu$ from 3 to 8, we see that in the $\theta = 0$ trace, there
are no $R_{xx}$ minima corresponding to the odd $\nu$, while there are strong
even-$\nu$ minima.  As the sample is tilted, the situation slowly reverses
itself, so
that at $\theta = 48.2^o$, there are no  minima
corresponding to the even $\nu$, but strong minima exist for the odd $\nu$.
This indicates that $\theta_1$ is near $48^o$.  This observation agrees with the
data of Smith {\it et al.}, which show the first coincidence to be roughly at
the same angle \cite{TPSmith87}.  However, Smith {\it et al.} reached the
conclusion that $|gm^*| = 1.52$ using Eq. 1 with $l = 1$ \cite{TPSmith87}.  This
conclusion is inconsistent with the remainder of our data.  If $l$ is taken to
be 1 for the first coincidence, then at $\theta = 0$
\begin{equation}
{|g|{\mu_B}B\over\hbar\omega_c} = l\cos\theta_1 = 0.7 \;.
\end{equation}
With this ratio, one would expect that at $\theta = 0$ the odd-$\nu$ $R_{xx}$
minima would be stronger than the even-$\nu$ minima.  Figure
\ref{lown} shows that the opposite is true.  Also, the angles of subsequent
coincidences are inconsistent with $l = 1$.  On the other hand, {\it all} of the
coincidences that we observe are consistent with $l = 3$ for $\theta_1$, $l = 4$
for $\theta_2$, etc.  This yields $|gm^*| = 4.1$.  Using $m^* =
0.46m_e$ \cite{Lay93}, we calculate that the Land{\'e} {\it g}-factor of
electrons confined to this AlAs QW is ${\pm}9.0$.  This {\it g}-factor is
consistent with the data of Smith {\it et al.} \cite{TPSmith87}, because
observation of the first coincidence alone cannot determine $|gm^*|$ to better
than the integer multiple $l$.

Other features of Fig. \ref{lown} are also consistent with $|g| = 9.0$.  Figure
\ref{fan}a is a plot of the energies of the Landau levels (LLs) for a tilt
experiment of an ideal, non-interacting 2DES with $|gm^*| = 4.1$ \cite{fannote}.
The spin-down (-up) levels are shown as solid (dotted) lines.  The coincidences
are marked with vertical lines and labelled in order.  When the Fermi energy
lies halfway between two of the LLs on the plot, the system is at an integer
$\nu$ and an $R_{xx}$ minimum is observed.   At a given angle, the energy gap
($\Delta_{\nu}$) between the LLs is the vertical distance between the LLs on the
plot.  Larger $\Delta_{\nu}$ are manifested as stronger $R_{xx}$ minima at that
$\nu$.  Qualitatively, all of the $R_{xx}$ minima in Fig. \ref{lown} have the
behavior described in Fig. \ref{fan}a.  For example,  Fig. \ref{fan}a predicts
that $\Delta_4$ (shaded for clarity) will be large at $\theta = 0$, disappear
completely at $\theta_1$, reach a maximum again at $\theta_2$, and remain
constant through all higher angles.  The $\nu = 4$ $R_{xx}$ minimum reflects
this behavior.

We also have similar tilt measurements of sample B gated to a density of $3.9
\times 10^{11}$ cm$^{-2}$, sample C at a density of $2.4 \times
10^{11}$ cm$^{-2}$, and sample D at a density of $3.6 \times 10^{11}$ cm$^{-2}$.
The data from all of the samples look similar, with all of the coincidences
happening at the same angles.  Since the quality is better at the higher
densities, more minima are observed at higher $\nu$, and they, too, follow the
behavior predicted by Fig. \ref{fan}a in the manner described above.  The
quality of the highest density data (from sample B) allows us make a more
precise measurement of the coincidence angles and therefore $|gm^*|$ than
would be possible with the data of Fig. \ref{lown} alone.  Data from sample B
are shown in Fig. \ref{fan}b:  the strengths of various $R_{xx}$ minima as they
evolve with $\theta$ are plotted.  This plot was made by subtracting a
linear background from the $R_{xx}$ vs. $B_{\perp}$ data, and plotting the new
$\Delta R_{xx}$ value for each integer $\nu$.  Since a particular $R_{xx}$
minimum is strongest when its corresponding $\Delta_{\nu}$ is largest, it is the
{\em minima} in Fig. \ref{fan}b that correspond to maxima in $\Delta_{\nu}$.  At
$\theta_1$ and $\theta_3$, the odd-$\nu$ curves in Fig. \ref{fan}b show minima,
and at $\theta_2$ the even-$\nu$ curves show minima \cite{nu5note}.  It is the
positions of the minima in Fig. \ref{fan}b that we used to calculate accurately
the angles of the coincidences, and therefore $|gm^*| = 4.1$, to within 4\%.  

The coincidence data provide a value for the ratio of the Zeeman
and cyclotron energies, i.e. $|gm^*|$, but not for the magnitude of these
energies individually.  The magnitude of $\Delta_{\nu}$ can be determined from
measurements of the activated behavior of the various $R_{xx}$ minima according
to $R_{xx} \propto \exp(-\Delta_{\nu}/2k_BT)$).  We have done such measurements
on sample B for the smaller fillings ($\nu = 1-3$) at various densities and
angles.  These measurements are consistent with
the Landau level diagram in Fig. \ref{fan}a, which indicates that $\Delta_1$ and
$\Delta_2$ should be $\hbar\omega_c$ at any $\theta$, and that $\Delta_3$ should
be $\hbar\omega_c$ for angles $\theta_1$ and above.  Shown in Fig. \ref{activ}a
are the measured $\Delta_{\nu}$ at various densities for $\nu = 1$ and $\nu = 2$
at $\theta = 0$ and for $\nu = 3$ at $\theta_1$.  The slope of the line fitted
to the points in Fig. \ref{activ}a is 3.4 K/T, in reasonable agreement with
$\hbar\omega_c$ which is expected to be 2.9 K/T.  The $\simeq15\%$ discrepancy
could come from the uncertainty in the mass measurement and also from the fact
that the measured $\Delta_{\nu}$ are reduced from the true
$\Delta_{\nu}$ by the disorder in the sample, which is expected to have a
smaller effect as the sample density is increased.  Therefore it is
reasonable that the slope of the line should be somewhat greater than the
expected slope for a system with no disorder.  The negative $y$-intercept of the
line in Fig. \ref{activ}a gives one estimate of the disorder in the sample: 14
K.  We get another estimate of roughly 9 K by examining the
$B_{\perp}$-dependence of the Shubnikov-de Haas oscillations \cite{Ando74a}.
The observation that the magnitude of the $y$-intercept (14 K) is larger than 9K
is also consistent with the disorder becoming less important as the density is
increased.  Finally, Fig. \ref{activ}b shows how some of the
$\Delta_{\nu}$ change as the sample is tilted.  The fact that $\Delta_1$ and
$\Delta_2$ do not rapidly increase as the sample is tilted is strong evidence
that neither $\Delta_1$ {\it nor} $\Delta_2$ are gaps of $g{\mu_B}B$.  Together,
all of these observations form a consistent picture that shows reasonable
agreement with the predictions of Fig. \ref{fan}a.   

The data we have presented so far all support the idea that this AlAs 2DES
behaves like the {\em non-interacting} Landau level diagram in Fig. \ref{fan}a.
There are some details, however, that are not explained by this picture.  
One is that at high densities, the $R_{xx}$ minima for $\nu$ up to 6 are
visible, although very weak, at angles at which they are expected to disappear
completely.  As Fig. \ref{fan}b shows, however, they
are at their weakest at the expected angles.  We do not understand this
unexpected anticrossing-like behavior.  The other is that, as the sample is
tilted, $\Delta_1$ and $\Delta_2$ fall with increasing $\theta$ (Fig.
\ref{activ}b) while Fig. \ref{fan}a indicates that they are expected to stay
constant at $\hbar\omega_c$.  However, the fact that both $\Delta_1$ and
$\Delta_2$ have the same behavior with $1/\cos\theta$ suggests that the same
effect is causing this deviation from the ideal behavior predicted by the Landau
level diagram.

The most interesting features of this 2DES are its apparent non-interacting
behavior and its unexpectedly large {\it g}-factor.    A constant
{\it g}-factor in this system is surprising given the results of previous
experiments which all show variations in $g$ that are well explained by
electron-electron interaction.  Ando and Uemura
proposed that this enhancement depends on the spin-population difference in the
2DES.  They conclude that the enhancement in $g$ for a given Landau level $N$
goes as $\sum_{N'}J^2_{NN'}(q)
(n_{N'\uparrow}-n_{N'\downarrow})$, where $n_{N'\uparrow}$ ($n_{N'\downarrow}$)
is the number of spin-up (-down)
electrons in the $N'$ Landau level \cite{Ando74}.  In the case of the Si
metal-oxide-semiconductor
structure, $J_{NN'}$ became negligible for $N'{\neq}N$.  Qualitatively, this is
true for all of the previously studied
systems \cite{Nicholas88,Englert82,Nicholas83,Weitz96,Shen96},  
because of the common feature they share:  for angles less than the first
coincidence angle there is only a spin-population 
difference when the Fermi energy lies within one Landau level (between the two
spin-split levels).  This is due to the fact that $|g|{\mu_B}B$ is
smaller than $\hbar\omega_c$ at $\theta = 0$.  These experiments were all
performed at angles near the first coincidence angle because with a smaller $g$,
the coincidences are at much higher angles, and features at the second
coincidence angle and beyond are not resolved.  In our AlAs QW sample, we have a
system in which $g{\mu_B}B$ (with $|g| = 9.0$), is significantly larger than
$\hbar\omega_c$ even at $\theta = 0$.  This not only leads to larger
spin-population differences, but also to a situation in which the Fermi energy
can {\em never} lie within one single Landau level.  Therefore it is some
different, and unknown, values of $J_{NN'}$ that are relevant to this system.
Under this picture, one hypothesis is that the enhancements due to
spin-population difference are not significant.  This would lead to the data
matching what would be expected of a non-interacting system of electrons.
However, this would not explain the magnitude of the {\it g}-factor.  The
expected bulk value from
theoretical calculations is 1.9 \cite{Roth59}, and the {\it g}-factor of
electrons in bulk Al$_{0.8}$Ga$_{0.2}$As has been measured by
electron-paramagnetic-resonance to be 1.96 \cite{Bottcher73}.  Also, van
Kesteren {\it et al.} have reported a value of $\simeq 1.9$ for electrons in
AlAs QWs based on optically detected magnetic resonance experiments on AlAs-GaAs
superlattices \cite{vanKesteren89}.  It could be that there is some other, still
unknown, electron
interaction-driven mechanism that is causing the enhancement seen here.  It is
also possible that the QW structure or some band structure effect is somehow
causing the enhancement over the bare value of 1.9.  If this is the case, it is
a very interesting development that warrants further study, because a better
understanding of the mechanism might allow one to use it to control the {\it
g}-factor independently of the other system parameters.

In summary, we have magnetoresistance and temperature dependence data revealing
that 2D electrons in a $150\AA$ QW behave as a non-interacting 2DES with a {\it
g}-factor of 9.0.  The coincidences observed in the magnetoresistance data
accurately determine $|gm^*| = 4.1$, and the activation energies agree with this
$|gm^*|$.  The magnitude of the {\it g}-factor is surprising because it remains
constant with $\nu$, and therefore appears to be enhanced by some unknown
mechanism other than the one that is observed in other 2DESs.

We would like to thank J. P. Lu, S. A. Lyon, and D. C. Tsui for useful
discussion and insight.  This work was funded by the NSF.

 \begin{figure}
 \caption{Magnetoresistance traces from a 2DES (density = $1.4 \times 10^{11}$
 cm$^{-2}$) in a 150 \AA-wide QW (sample A) at various angles of tilt.}
 \label{lown}
 \end{figure}

 \begin{figure}
 \caption{a: Diagram of the Landau level energies for a tilt experiment in a
 non-interacting 2DES with $|gm^*| = 4.1$.  The solid (dashed) lines correspond
 to spin-up (-down) Landau levels.  b: $\Delta R_{xx}$ points as a measure of
 the relative strengths of the $R_{xx}$ minima.  The $\Delta R_{xx}$ were
 calculated by subracting a linear background from the $R_{xx}$ vs. $B_{\perp}$
 data.}
 \label{fan}
 \end{figure}

 \begin{figure}
 \caption{a:  Activation energies from sample B.  The activation
 energy for $\nu = 2$ was measured at various densities.  b:  Activation
 energies at various $\theta$ measured in sample B.}
 \label{activ}
 \end{figure}


\begin{thebibliography}{10}
\bibitem{fannote}
It should be noted that only the product $gm^*$ has significance in this
  diagram. Any system with a given $gm^*$ will have the same Landau level vs.
  tilt diagram for the ideal, non-interacting case, regardless of what the
  individual values of $g$ and $m^*$ happen to be.
\bibitem{nu5note}
Note that from the Landau level diagram, $\Delta_5$ is expected to remain
  strong after the third coincidence and Fig. \ref{fan}b shows this.
\end{thebibliography}
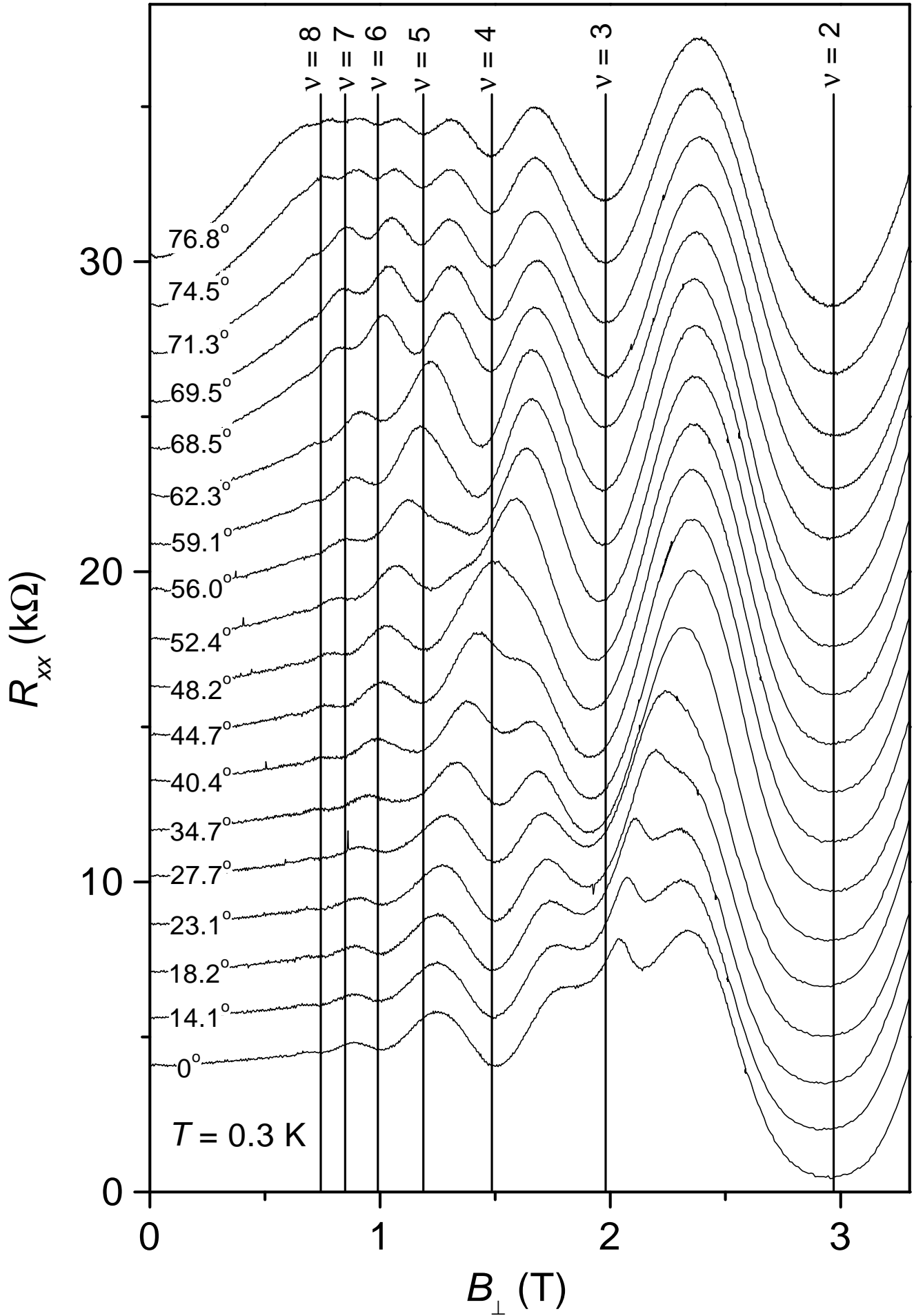
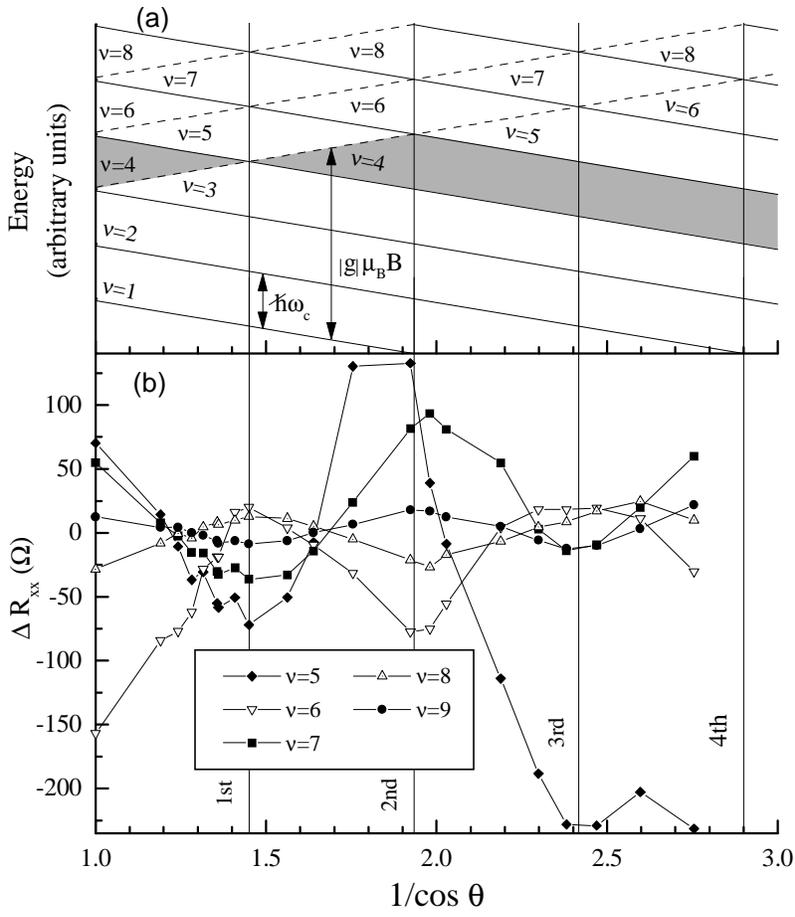
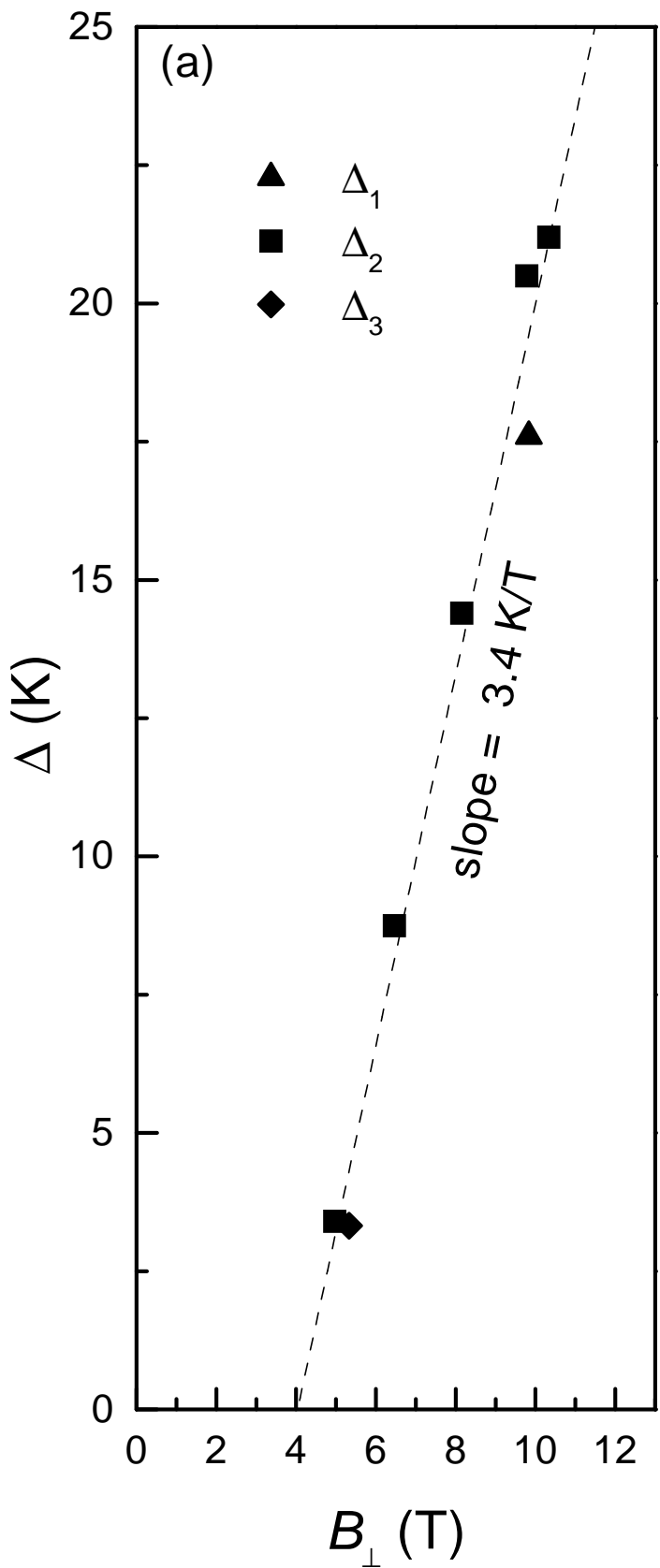
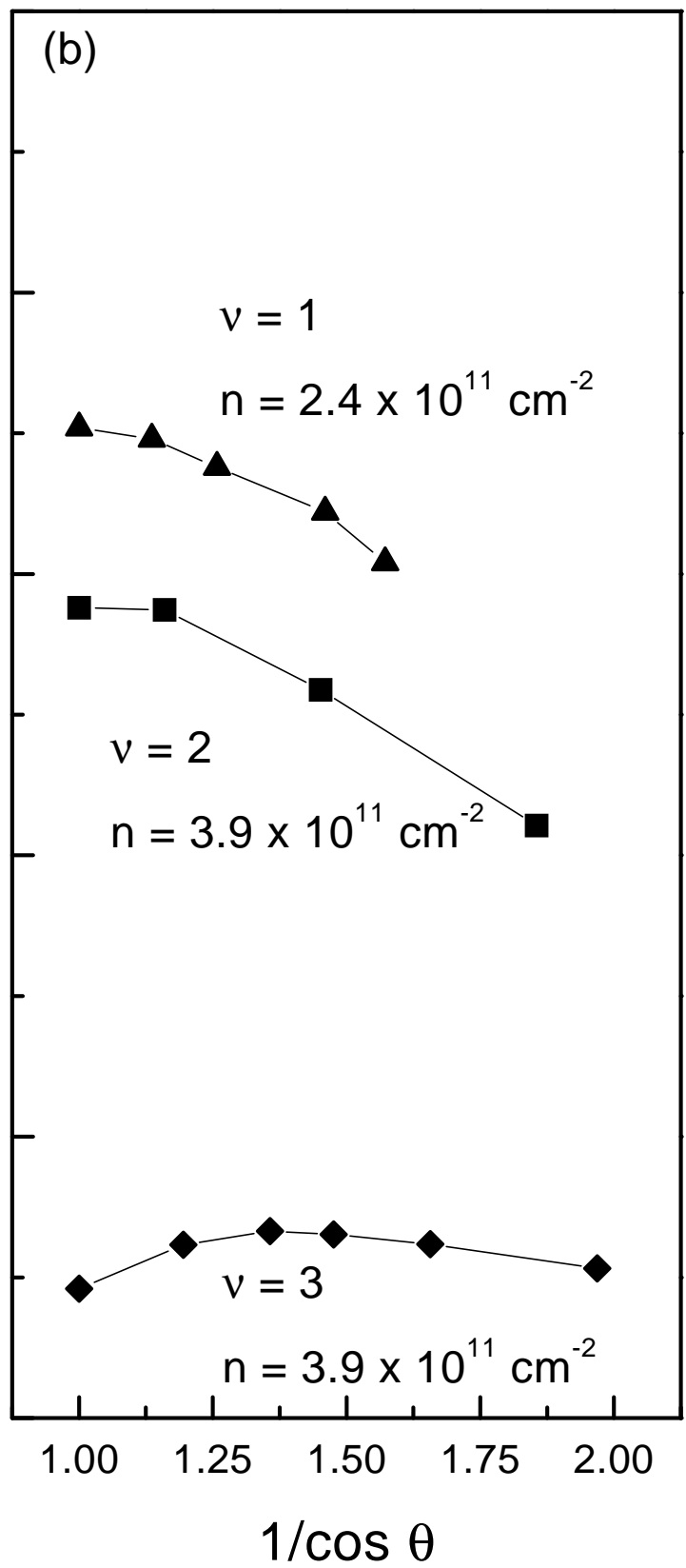
\end{document}